\documentclass[
reprint,
amsmath, amssymb,
pra,
aps,
superscriptaddress, twocolumn,nofootinbib,groupedaddress,
longbibliography,showkeys
]{revtex4-1}

\usepackage{graphicx}
\usepackage{subfigure}
\usepackage{hyperref}
\usepackage{bm}
\usepackage{float}
\usepackage{color}
\usepackage{amsmath}
\usepackage{braket} 
\usepackage{ulem}

\hypersetup{colorlinks=true, citecolor=blue, urlcolor=blue, linkcolor=blue}
\begin{document}
\title{Enhanced Microwave Sensing with Dissipative Continuous Time Crystals}
\author{Yunlong Xue$^{1, 2}$}
\author{Zhengyang Bai$^{1}$}
\email{zhybai@nju.edu.cn}
\author{Yu-Qiang Ma$^{1, 3}$}
\email{myqiang@nju.edu.cn}
\affiliation{
	 $^1$National Laboratory of Solid State Microstructures and School of Physics,
	Collaborative Innovation Center of Advanced Microstructures, Nanjing University, Nanjing 210093, China\\
		$^2$State Key Laboratory of Precision Spectroscopy, East China Normal University, Shanghai 200062, China\\
    $^3$Hefei National Laboratory, Hefei 230088, China}
\date{\today}


\begin{abstract}
A dissipative time crystal is an emergent phase in driven-dissipative quantum many-body systems, characterized by sustained oscillations that break time-translation symmetry spontaneously.
Here, we explore nonequilibrium phase transitions in a dissipative Rydberg system driven by a microwave (MW) field and demonstrate their critical sensitivity to high-precision MW sensing. Distinct dynamical regimes are identified, including monostable, bistable, and oscillatory phases under mean-field coupling.
Unlike single-particle detection—where the beating signal decays linearly with MW field strength—the time crystalline phase exhibits high sensitivity to MW perturbations, with rapid, discontinuous frequency switching near the monostable-oscillatory boundary. The abrupt transition is rooted in spontaneous symmetry breaking in time and is fundamentally insensitive to the background noise.
On this basis, a minimum detectable MW field strength on the order of $1\,\mathrm{nV/cm}$ is achieved by leveraging this sensitivity.
Our results establish a framework for controlling time crystalline phases with external fields and advance MW sensing through many-body effects.
\end{abstract}


\maketitle

\textit{Introduction}.--- 
Neutral atom in high-lying electronic Rydberg states (principal quantum number $n\gg 1$) is a powerful platform with promising potential applications in
quantum optics, quantum simulation, and non-equilibrium dynamics \cite{2d29,2d31,2d34}. This is rooted in their ability to provide both strong, long-range atomic interactions and high controllability through laser excitation, which satisfies the prerequisite for studying complex many-body physics~\cite{bernien2017scars}. For closed quantum systems, a plethora of studies have been made by using Rydberg atoms (especially neutral-atom arrays) in probing exotic quantum phenomena,  such as  quench dynamics~\cite{quench_GS_2018,quench_chen_2025,bernien2017scars} , thermalization~\cite{thermalization_khemani_2019,thermalization_kim_2018,thermalization_zhao_2025,thermalization_michailidis_2020,thermalization_he_2025}, ergodicity breaking~\cite{serbyn2021quantum, bluvstein_controlling_2021} and information scrambling~\cite{li2024Observation, liang2024Observation}.

At finite temperatures, dissipation often becomes non-negligible due to coupling with the local environment. Dissipation generally destroys quantum many-body coherence and entanglement, driving the system to stationary states within its relaxation timescale. On the other hand, the synergy between Rydberg interactions and dissipation gives rise to exotic non-equilibrium phenomena, such as collective quantum jumps
\citep{2d34, xiang2025Switching}, and self-organized criticality~\cite{Marcuzzi2016,Ding2020a,helmrich2020signatures}, optical bistability~\cite{Antiferromagnetic_Lee_2011,carr2013Nonequilibrium, weller2016}, and collective oscillation~\cite{Wadenpfuhl2023,YL,Ding2024}.

Recent experiments have demonstrated that  nonlinear synchronization is enhanced in a thermal rubidium gas involving two Rydberg states, forming a dissipative continuous time crystal (CTC)~\cite{YL}. This phenomenon arises from the competition between the two Rydberg states, which induces interaction-driven limit cycles. These limit cycles can break time-translation symmetry, causing the system to bifurcate from a uniform phase into an oscillatory phase.
The persistent oscillations in this phase critically depend on the interplay between the two Rydberg states, making it potentially susceptible to disruption by external fields coupling their transition. This susceptibility, in turn, suggests a novel approach for detecting external electromagnetic fields [e.g., microwave (MW) or terahertz fields] using CTC phases~\cite{sedlacek2012Microwave,jing2020,ding2022enhanced,li2024Room,tu2024Approaching}.

In this work, we examine this problem by studying 
nonequilibrium dynamics in a MW-coupled Rydberg many-body system. Under mean-field (MF) coupling, we identify three distinct phases in the phase diagram: monostable, bistable, and oscillatory phases. Crucially, we find that the oscillatory regime is exceptionally sensitive to variations in the MW field. This sensitivity manifests as a discontinuous, rapid switching between non-equilibrium states characterized by high- and zero- oscillation frequencies with increasing MW power. This is the key finding of this work.

Furthermore, numerical simulations that include Doppler effects confirm that this  frequency switching near criticality remains robust under thermal averaging. Building on this, we propose a high-precision microwave measurement protocol based on thermal Rydberg gases. By leveraging this sensitivity, the proposed scheme achieves a minimum detectable electric field strength on the order of 
$1\,\mathrm{nV/cm}$, underscoring its potential for MW-field detection.

Of particular relevance, Ding et al. demonstrated enhanced MW sensitivity using the steep slope at the bistable critical point in Rydberg atomic gases~\cite{ding2022enhanced, whitlock2022Many}. Beyond bistability, the oscillatory regime offers a promising yet unexplored avenue. Crucially, the rapid switching between the time-crystalline and stationary phases is readily probed with a standard spectrum analyzer~\cite{jing2020,jiao2025Arbitrary}, making it a candidate for MW detection. 
\begin{figure}[htb]
	\includegraphics[width=0.5\textwidth]{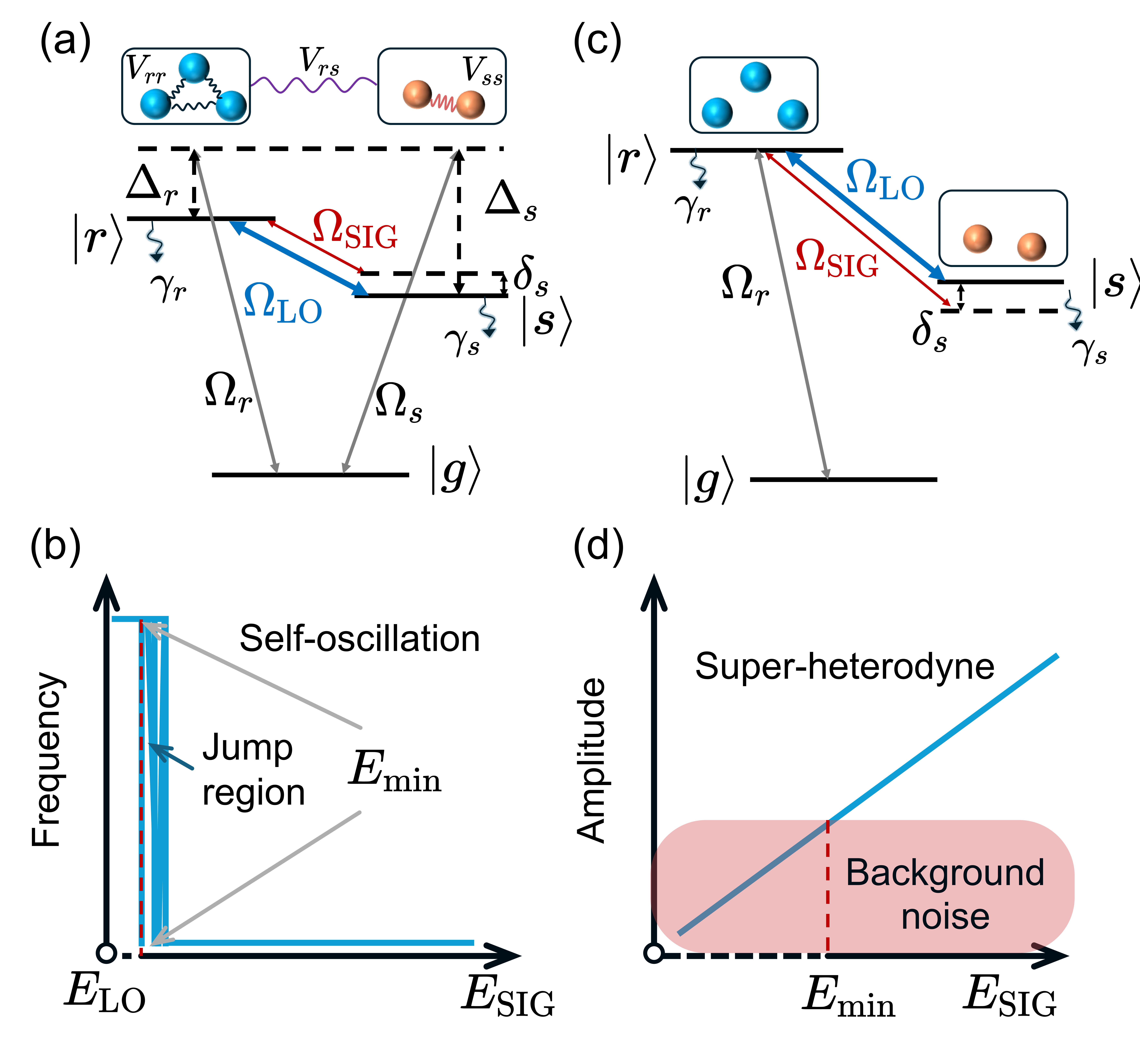}
	\caption{{\bf Basic principle of the CTC-based MW receiver.} (a) Energy-level diagram. The drive fields $\Omega_s$ ($\Omega_r$) directly couple $|g\rangle$ to $|s\rangle$ ($|g\rangle$ to $|r\rangle$), while the local MW field $\Omega_{\mathrm{LO}}$ resonantly couples $|r\rangle$ and $|s\rangle$. A weak MW signal field $E_{\mathrm{SIG}}$ (Rabi frequency $\Omega_{\mathrm{SIG}}$) has detuning $\delta_s$ and phase $\phi_s$ relative to the local field.
Here, the controllable local field $\Omega_{\mathrm{LO}}$ locks the system near the onset of the phase transition, significantly enhancing its sensitivity to the weak signal field $E_{\mathrm{SIG}}$.
(b) Oscillation frequency of the CTC phase versus the signal field strength $E_{\mathrm{SIG}}$.
(c) Energy-level diagram of the atomic superheterodyne MW receiver. The local field $\Omega_{\mathrm{LO}}$ resonantly couples states $|r\rangle$ and $|s\rangle$, while the signal field $\Omega_{\mathrm{SIG}}$ drives the transition to $|s\rangle$ with a detuning $\Delta_s$, generating a beat-oscillation signal.
(d) Measured signal amplitude versus the signal-field strength $E_{\mathrm{SIG}}$. The ultimate sensitivity limit for MW detection is determined by the background noise.}
	\label{fig:1}
\end{figure}

\textit{Model}.---
We consider a Rydberg ensemble with three electronic levels, including a ground state $|g\rangle$ and two Rydberg states $|r\rangle$ and $|s\rangle$,  as depicted in Fig.~\ref{fig:1}(a).
Each atom is continuously laser-excited  from the ground state $\ket{g}$ to Rydberg state $\ket{r}$ ($\ket{s}$) with Rabi frequency $\Omega_r$ ($\Omega_s$) and detuning $\Delta_r$ ($\Delta_s$) from resonance.  
Under the rotating-wave approximation, the atom-light coupling Hamiltonian of the system is given by, 
$\hat{\mathcal{H}}_{\rm al}  = -\sum_{\alpha=r, s}\Delta_\alpha\hat{\sigma}_{\alpha\alpha}(\mathbf{r},t)+\frac{1}{2}[\Omega_{r}\hat{\sigma}_{rg}(\textbf{r},t)
+\Omega_{s}\hat{\sigma}_{sg}(\textbf{r},t)
		+\text{H.c.}]$.
Here, $\hat{\sigma}_{\alpha\beta}$ are transition operators $(\alpha,\beta = g, r, s)$.

Once excited, the Rydberg atoms interact via strong and long-range interactions. Typically, the interactions $V_{\alpha\beta}(\mathbf{r}-\mathbf{r}^{\prime}) = C_{\alpha\beta}^{(6)}/|\mathbf{r}-\mathbf{r}^{\prime}|^{6}$ are described with the van der Waals interaction with dispersion coefficient  $C_{\alpha\beta}^{(6)}$ ($\alpha, \beta = r, s$) between the Rydberg atoms located at the positions $\mathbf{r}$ and $\mathbf{r}^{\prime}$, respectively~\cite{YL, ding2022enhanced}. Given the presence of two Rydberg states, the two-body interaction is governed by the Hamiltonian,
\begin{equation}
	\begin{aligned}
		H_{\rm int} =&-\mathcal{N}\sum_{\alpha, \beta}\int d^{3}\mathbf{r}^{\prime}\hat{\sigma}_{\alpha\alpha}\left(\mathbf{r}^{\prime},t\right)V_{\alpha\beta}(\mathbf{r}-\mathbf{r}^{\prime})\hat{\sigma}_{\beta\beta}\left(\mathbf{r},t\right),
	\end{aligned}
\end{equation}
To implement a CTC-based MW receiver, we couple two Rydberg states via MW fields. The system is governed by the Hamiltonian, $\hat{\mathcal{H}}_{\rm am}  = \frac{1}{2}\Omega_{m}\hat{\sigma}_{rs}(\textbf{r},t)+\text{H.c.}$, where $\Omega_m$ is the MW Rabi frequency.

Meanwhile, the coherent laser coupling and strong long-range interactions in Rydberg states compete with dissipative processes (e.g., thermal collisional effects and electronic state decay). This dissipation is characterized by the decay rates $\gamma_r$ and $\gamma_s$ of the Rydberg states.
To incorporate dissipation effects, the evolution of the density matrix $\hat{\rho}$ is governed by the Lindblad equation,
\begin{equation}\label{eq;5}
	\frac{\partial \hat{\rho}}{\partial t} = -i \left[ \hat{H}, \hat{\rho} \right] + \mathcal{D} \left( \hat{\rho} \right),
\end{equation}
Here, the total system Hamiltonian is given by $\hat{H}=\mathcal{N}\int d\mathbf{r}^3 (\hat{\mathcal{H}}_{\rm al}+\hat{\mathcal{H}}_{\rm am}+\hat{\mathcal{H}}_{\rm{int}})$, where $\mathcal{N}$ denotes the atomic density. 
The Lindblad term $\mathcal{D}(\hat{\rho})$ is defined as
$\mathcal{D} (\hat{\rho} ) = \mathcal{N}\sum_{\alpha=r, s} \gamma_{\alpha}\int d^{3}{\mathbf r} ( \hat{L}_\alpha \hat{\rho} \hat{L}_\alpha^\dagger - \frac{1}{2} \{ \hat{L}_\alpha^\dagger \hat{L}_\alpha, \hat{\rho} \} )$,
with Lindblad operators $\hat{L}_\alpha = \hat{\sigma}_{g\alpha}(\mathbf{r}, t)$ corresponding to the decay rates $\gamma_\alpha$.
The first term on the right-hand side of Eq.~(\ref{eq;5}) describes the unitary evolution of $\hat{\rho}$ governed by $\hat{H}$.
The Lindblad term accounts for dissipative processes from the Rydberg states $|r\rangle$ and $|s\rangle$ to the ground state $|g\rangle$.
Without loss of generality, we assume identical decay rates $\gamma_r = \gamma_s = \gamma$ for adjacent Rydberg states, and hereafter express all quantities in units of $\gamma$.

The interplay between strong Rydberg interactions and dissipation results in exotic CTC phases that have recently been demonstrated experimentally in several works~\cite{Wadenpfuhl2023,YL,Ding2024,jiao2025Observation}. Here, we aim to leverage CTC phases to enhance MW sensing, with the underlying principle described as follows. 
A strong local-oscillator MW field, described by the Rabi frequency $\Omega_{\mathrm{LO}}$ is applied resonantly to control the system through nonequilibrium phase transitions.  
Detection is performed by introducing a weak signal field of the form $\Omega_{\mathrm{SIG}} =\tilde{\Omega}_{\mathrm{SIG}}
\exp(-2\pi i\delta_s t + \phi_s)$,
which couples the same Rydberg states under the condition $\Omega_{\mathrm{SIG}} \ll \Omega_{\mathrm{LO}}$ [Fig.~\ref{fig:1}(a)]. 
Here, the frequency detuning $\delta_s$ and phase $\phi_s$ are defined relative to the local field, and the condition $2\pi|\delta_s| < \Gamma$ needs to be satisfied, where $\Gamma$ denotes the system's linewidth. The Rabi frequency of the MW field is therefore given by $\Omega_m=\Omega_{\mathrm{LO}}+\Omega_{\mathrm{SIG}}$.

The proposed MW receiver shares a structural analogy with an atomic superheterodyne receiver [Fig.~\ref{fig:1}(c)]. In a superheterodyne architecture, the interference between the local and signal fields produces a beat signal whose amplitude scales linearly with the MW field strength [Fig.~\ref{fig:1}(d)]. As a result, the ultimate sensitivity limit for MW detection is determined by the background noise—comprising laser, electrical, and atomic contributions—and is reached when the signal‑to‑noise ratio approaches unity~\cite{jing2020}.

In the CTC-based MW receiver, the system is operated near the oscillatory–monostable boundary, tuned via the local field $\Omega_{\rm LO}$. Prior to crossing the phase transition point, spontaneous breaking of time-translation symmetry occurs, leading to self-oscillations analogous to a self-beating signal. Once the signal field $E_{\rm SIG}$ exceeds a threshold, these oscillations abruptly cease, accompanied by rapid and discontinuous frequency switching (the detailed calculation is shown in the next section). This abrupt change originates from spontaneous symmetry breaking in the system and is essentially independent of the background noise [Fig.~\ref{fig:1}(b)]. As a result, enhanced MW detection precision can be anticipated near the critical point, as will be elaborated in the following analysis.

\textit{Microwave-driven nonequilibrium  phase transition}.---
To this end, we first investigate the nonequilibrium phase transition driven solely by local MW fields. In this context, the Rabi frequency of the MW field is denoted as $\Omega_m=\Omega_{\mathrm{LO}}$. 
The three-state system with basis states $|\alpha\rangle$ ($\alpha=g, r, s$) can be mapped onto an effective spin-1 system. The general density matrix for this system is then given by $\hat{\rho}_l=\sum_{\alpha,\beta=g, r, s} c_{\alpha\beta}|\alpha\rangle\langle\beta|$.
For a many-particle system, 
the dimension of the density matrix grows as $8^{N}$ in Liouville space. The computational complexity prevents us from numerically solving the many-body problems when $N > 10$ on typical computers. 
Due to the dissipation and long-range interactions in Rydberg ensemble, we can employ the MF theory to  characterize the dynamical behavior of this system~\cite{lee2011Antiferromagnetic, carr2013Nonequilibrium}. In the MF approximation, the many-body density matrix $\hat{\rho}$ is decoupled into individual ones through a MF ansatz $\hat{\rho}=\prod_{l}{\otimes\hat{\rho}_l}$. This decoupling essentially ignores spatial correlations between different sites~\cite{diehl2010Dynamical,YL}. 
Based on the Lindblad master equation (\ref{eq;5}), we obtain the optical Bloch equations of density-matrix elements $\rho_{\alpha\beta}({\bf r},t)\equiv\langle {\hat \sigma}_{\alpha\beta}({\bf r},t)\rangle$ [see Supplementary Material ({\bf SM}) for details].
Due to the Rydberg interactions, eight components in the MF equations are coupled nonlinearly  (because nondiagonal elements are complex).
\begin{figure}[htb]
	\centering
	\includegraphics[width=0.5\textwidth]{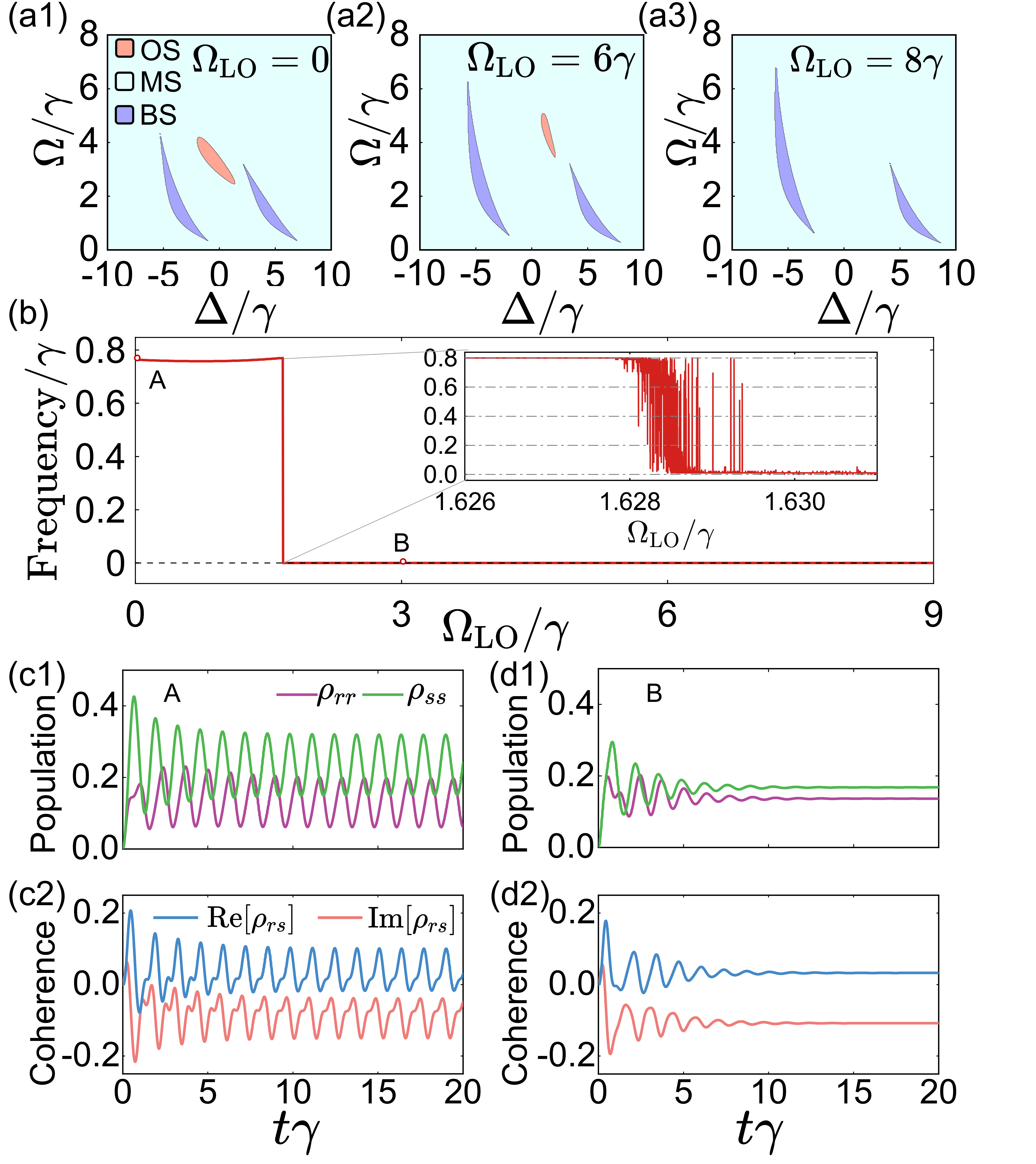}
	\caption{{\bf  Mean-field phase diagram and analysis of the CTC phase.} (a1)-(a3) We sketch the  phase diagram in $\Omega-\Delta$ space. The phase diagram is occupied by  the  MS,  BS, and  OS states. The panels (a1),(a2) and (a3) correspond to  the phase diagram where MW field $\Omega_{\rm LO}=0$, $6\gamma$, and $8\gamma$, respectively. (b) The Fourier analysis for the dynamics of population $\rho_{gg}$ as a function of   $\Omega_{\rm LO}$ is  obtained  with $\Omega=3.3\gamma$ and $\Delta=-0.6\gamma$. Near the oscillatory–monostable boundary, the oscillation frequency exhibits fast fluctuations as it drops rapidly from a high value to nearly zero.
		The dynamics corresponding to points A and B are presented in panels (c1)-(c2) and (d1)-(d2), respectively. Panels (c1)-(c2) plot the Rydberg populations $\rho_{\alpha\alpha}$ $(\alpha=r, s)$ and the coherence $\rho_{rs}$ versus time $t$ at $\Omega_{\rm LO}=0$. Panels (d1)-(d2) show that increasing $\Omega_{\rm LO}$ to $3\gamma$ destroys the sustained oscillation.  Throughout, the system parameters are fixed: $\gamma_r = \gamma_s = \gamma$ and $V_{rr} = V_{ss} = V_{rs} = 10\gamma$.}
	\label{fig:2}
\end{figure}

The fixed points of the optical Bloch equations are found by setting $\dot{\rho}_{\alpha\beta}=0$. We further employ the Lyapunov stability criterion to quantify the linear stability of the fixed points by calculating eigenvalues of the Jacobian matrix of the MF equations~\cite{Strogatz2015Nonlinear} . 
To characterize the phases, we will use Rydberg population $\rho_{rr} (\rho_{ss})$  as order parameters. Based on the number of fixed points and their stability,  the phase diagram in the $\Delta-\Omega$ plane is partitioned into three distinct regimes: monostable (MS), bistable (BS), and oscillatory (OS) states [see Fig.~\ref{fig:2}(a1)-(a3)].  The latter two correspond to nontrivial many-body phases~\cite{lee2011Antiferromagnetic, ddsbzy} [see {\bf SM} for details].

In our simulation, we vary the detuning $\Delta$ (with \(\Delta_{s} = \Delta\) and \(\Delta_{r} = \Delta - 8\gamma\)) and the Rabi frequency $\Omega$ (with \(\Omega_{r} = \Omega_{s} = \Omega\)) with different the local MW field $\Omega_{\rm LO}$.  
As shown in Fig.~\ref{fig:2}(a1), the MF phase diagram comprises all three phases in the absence of a MW field (e.g., when $\Omega_{\rm LO} = 0$).
In the monostable regime (cyan area), the dissipation (e.g., the decay $\gamma$) is dominant  and  relaxes the system to a  stationary state within  the atomic lifetime $T\sim1/\gamma$.   In the bistable regime (blue area), the nonlinearities (contributed from Rydberg interactions) drive a saddle-node bifurcation, and the system exhibits a first-order dissipative phase transition between states with high and low Rydberg populations~\cite{carr2013Nonequilibrium, xiang2025Switching}.  

As the detuning $\Delta$ approaches resonance, the competition between the two Rydberg states causes the system to bifurcate from a monostable regime into an oscillatory phase. 
This occurs due to the emergence of interaction-induced LCs that spontaneously break time-translation symmetry through a Hopf bifurcation~\cite{cross1993pattern,lee2011Antiferromagnetic}.  Having identified the regime for oscillatory states, we find that sustained oscillations can arise from the interplay between two Rydberg states [Fig.~\ref{fig:2}(c1)-(c2)]. This behavior constitutes a dissipative CTC~\cite{YL}. 

In the presence of local MW fields [Fig.~\ref{fig:2}(a2)], we observe that the bistable regime in the phase diagram remains relatively stable, without significant broadening or frequency shift. In contrast, the oscillatory regime is highly sensitive to the MW field. As the local field $\Omega_{\rm LO}$ increases from $0$ to $8\gamma$, the induced stark shift causes a distinct, blue-detuned oscillatory regime to emerge. This regime is characterized by a progressive narrowing until its eventual disappearance [Fig.~\ref{fig:2}(a3)].

As discussed above, the CTC phase emerges due to the interplay between two Rydberg states. That also leads to  an oscillating atomic coherence $\rho_{rs}$ between two Rydberg states [Fig.~\ref{fig:2}(c2)]. 
As demonstrated in Fig.~\ref{fig:2}(d2), the local MW field effectively modulates the atomic coherence $\rho_{rs}$ between two Rydberg states, leading to suppression of coherent oscillations and subsequent relaxation of the system to a stationary state.

To understand the dissipative phase transition controlled by local fields, 
we map the phase diagram as a function of the local field $\Omega_{{\rm LO}}$ through Fourier analysis of the persistent oscillations in $\rho_{gg}$. The system parameters are fixed at $\Delta=-0.6 \gamma$ and $\Omega = 3.3 \gamma$.
At the boundary between the monostable and oscillatory regimes, a rapid switching emerges between two distinct states. This transition is characterized by a sharp jump from a state with high oscillation frequency to one with nearly zero frequency as the local field $\Omega_{\rm LO}$ is continuously increased [Fig.~\ref{fig:2}(b)]. 
Due to the high-sensitivity of the oscillatory regime to the MW field, this phenomenon can provide a novel many-body approach for MW sensing based on Rydberg atoms~\cite{jing2020,ding2022enhanced}. These findings indicate that by operating the system near the oscillatory–monostable boundary via the local field, we can achieve high-sensitivity detection of signal MW fields. The following section presents a detailed analysis of this CTC-based MW receiver implemented in thermal atomic gases.

\textit{Protocol of a CTC-based MW receiver}.---
\begin{figure}[b]
	\includegraphics[width=0.5 \textwidth]{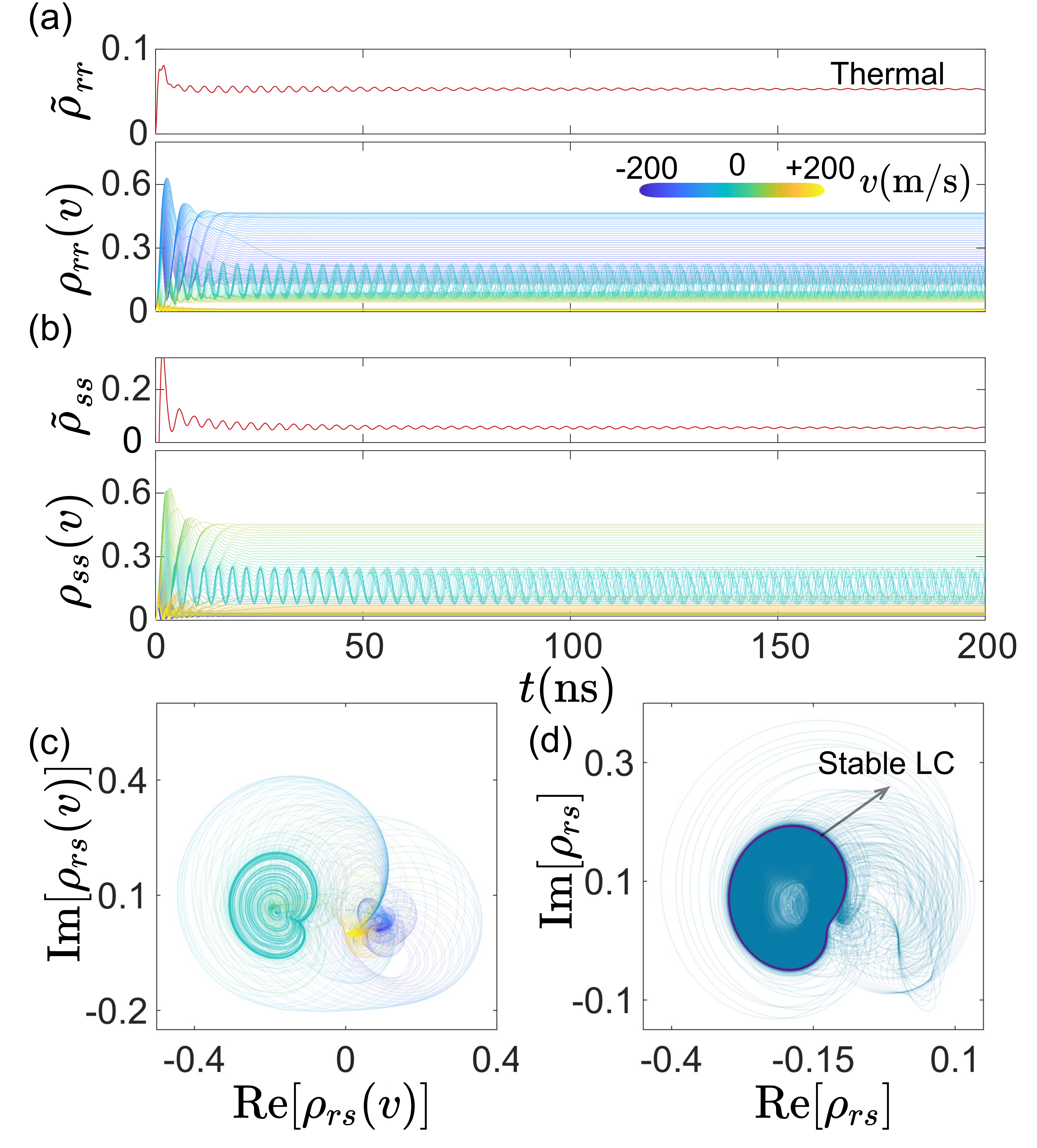}
	\caption{{\bf Dynamical evolution of Rydberg populations across velocity classes.} 
		(a) [(b)] Time evolution of \(\rho_{rr}\) (\(\rho_{ss}\)) 
	for different velocity classes ranging from \(-200\) m/s to \(200\) m/s (lower panel), along with their thermally averaged results (upper panel). (c) Phase-space evolution trajectories for atoms with different velocities. Panels (a), (b), and (c) share the same color mapping, and all atoms are initialized in the ground state. (d) Phase space trajectories for atoms with a fixed velocity ($v=0$) but different initial population distributions. In the simulation, parameters are \(T = 300\)K, \(\Omega = 3\gamma\), \(\Omega_{\rm LO} =\gamma\), \(V_{rr} = V_{ss} = V_{rs} = 10\gamma\),  and \(\Delta = 0\).}
	\label{fig:3}
\end{figure}
For the practical application of the Rydberg MW receiver, we extend the above study of CTCs to a thermal atomic system.
In a thermal atomic vapor, the high atomic density gives rise to strong Rydberg interactions that dominate the system dynamics~\cite{carr2013Nonequilibrium}. The random thermal motion of atoms in a Rydberg vapor results in a velocity distribution that induces inhomogeneous Doppler broadening~\cite{hu2025Thermal}. The Doppler effect plays a critical role in shaping atomic dynamics.

\begin{figure*}
	\includegraphics[width=1 \textwidth]{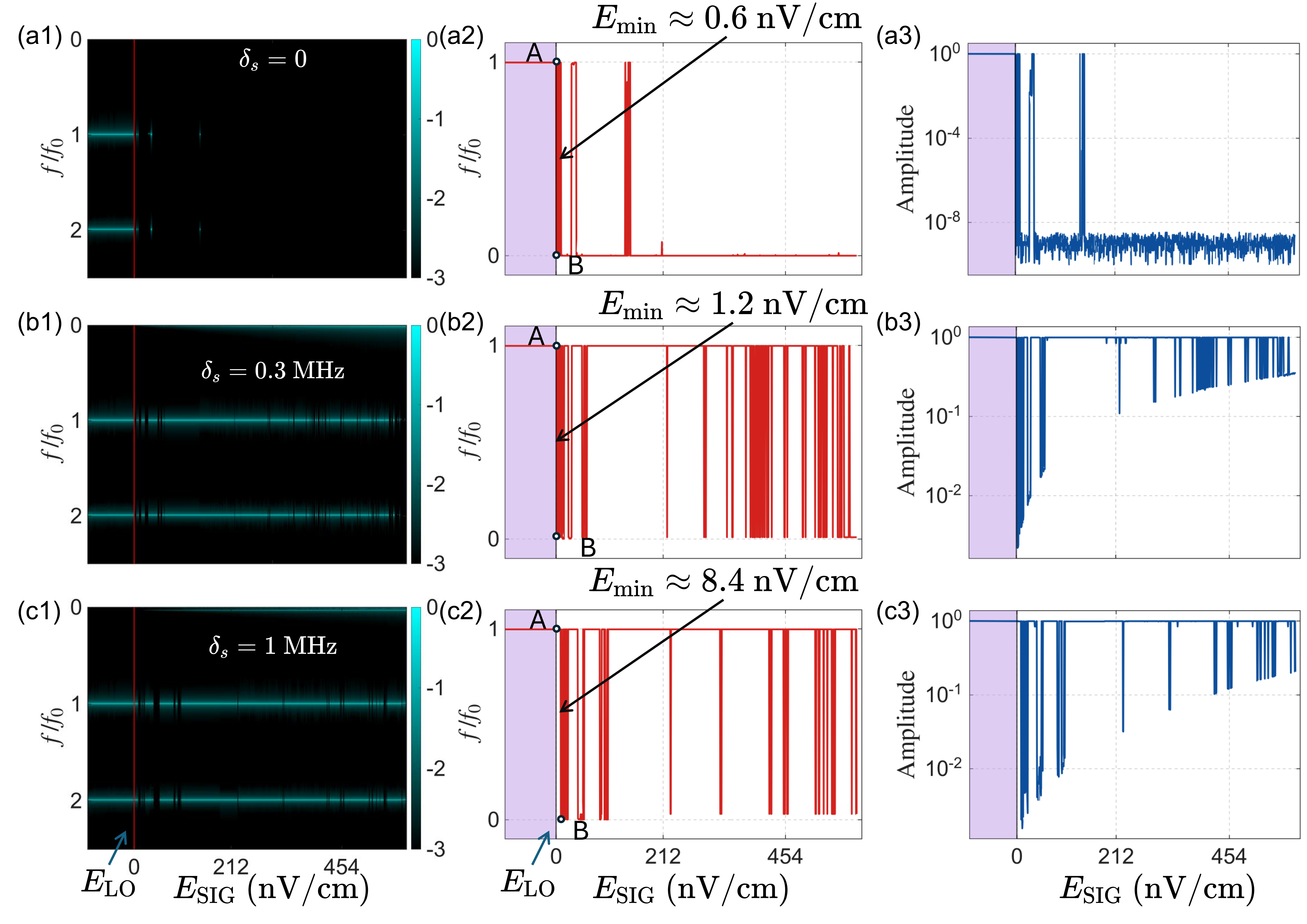}
\caption{\textbf{MW sensing via a dissipative CTC.}
(a1--c1) Oscillation frequency versus the signal-field strength $E_{\mathrm{SIG}}$ at a fixed local field $E_{\mathrm{LO}} = 7.85 \times 10^{-3} \, \mathrm{V/cm}$. The signal detuning $\delta_s$ is: (a1)-(a3) 0 (resonant), (b1)-(b3) $0.3 \, \mathrm{MHz}$, and (c1)-(c3) $1 \, \mathrm{MHz}$. Color represents the normalized oscillation amplitude (log scale). The red vertical line marks the locking point under fixed $E_{\mathrm{LO}}$, with $E_{\mathrm{SIG}}$ increasing to its right. To avoid overcrowding the figure, the beating frequency is not displayed. 
(a2--c2) Corresponding normalized fundamental frequency $f/f_0$ as a function of $E_{\mathrm{SIG}}$. The pink band indicates the accessible $E_{\mathrm{LO}}$ range, and the black line denotes the actual locking point used in the simulation, corresponding to the red line in (a1--c1).
(a3--c3) Amplitude of the fundamental oscillation versus $E_{\mathrm{SIG}}$.
The abrupt transitions in both frequency and amplitude enable the detection of MW fields.}
\label{fig:4}
\end{figure*}
This simulation accounts for atomic velocity effects by incorporating Doppler averaging for each transition. The Doppler broadening is included in the Hamiltonian through the detuning,
$\Delta_\alpha\rightarrow \Delta_\alpha+\mathbf{k}_\alpha\cdot\vec{v}$ $(\alpha = r, s)$. The effective density matrix is then obtained by velocity averaging, $\tilde{\rho}_{\alpha\beta}=\int dv f(v) \rho_{\alpha\beta}(v)$, where $f(v)$ represents the Maxwell–Boltzmann velocity distribution, $f(v)=1/(\sqrt{\pi}v_T){\rm exp}[-(v/v_T)^2]$, and $v_T=\sqrt{2k_BT/M}$ is the most probable atomic speed at temperature $T$ ($M$ mass of atoms).

As illustrated in Fig.~\ref{fig:3}(a)[(b)], we analyze the dynamical behavior of the system across different velocity classes, with special emphasis on Doppler broadening effects. It is found that dynamics of atoms with different velocity classes are largely different during the initial transient period. At a later stage ($t> 20$ ns),  we observe that a substantial portion of atoms attains a steady-state population, becoming dynamically inactive. In contrast, a significant fraction of velocity classes remains dynamically active, exhibiting persistent temporal oscillations.

Notably, despite the initial divergence in their amplitudes and frequencies, the oscillations of different velocity classes ultimately align in phase, demonstrating a transition toward collective behavior. In the upper panels of Fig.~\ref{fig:3}(a)–(b), we plot the Doppler-averaged Rydberg populations, $\tilde{\rho}_{ss}$ and $\tilde{\rho}_{rr}$. After an initial transient period, the mean population enters a sustained oscillatory phase. The oscillation frequencies of different velocity classes are locked and become centered at a single frequency. 
Such collective oscillation resembles the CTC phase in the thermal Rydberg gases~\cite{Wadenpfuhl2023,YL,Ding2024}.

We next examine the evolution of atomic coherence across different velocity classes in phase space. Our results reveal that particles within the oscillatory interval organize into a series of concentric limit cycles, sustaining persistent oscillations without decay [Fig.~\ref{fig:3}(c)]. Additionally, we explore the impact of initial conditions on the system's dynamics. As depicted in Fig.~\ref{fig:3}(d), different initial population distributions generate distinct phase-space trajectories with varying phases. This behavior is consistent with recent findings in Ref.~\cite{Wadenpfuhl2023}, that serves as a key signature distinguishing time crystals from other systems. Remarkably, all trajectories eventually stabilize onto the same limit cycle, demonstrating the robustness of time-crystalline oscillations.

We then study the dynamical response of the dissipative CTC to a MW signal field in a thermal gas. 
To exploit the high sensitivity near the oscillatory--monostable boundary, we first lock the local field close to the critical point. 
In this configuration, the system operates in the CTC regime and exhibits a stable oscillation. 
After preparing the system in the initial CTC state, we apply a weak signal field to perturb it. 
Owing to the dynamical instability near the critical point, the CTC state can be disrupted under weak signal-field modulation, causing the oscillation frequency to drop to zero. 
Here, we define the minimum detectable field strength as the signal intensity required to drive the oscillation frequency from its CTC value down to the point where the oscillatory signal first vanishes.

As an example, we choose the highly excited Rydberg states of rubidium atoms $|r\rangle=|90S_{1/2}\rangle$  and $|s\rangle=|90P_{3/2}\rangle$, respectively. The other parameters are chosen with detunings $\Delta_{r} = \Delta_{s} = 6\gamma$ and  Rabi frequencies $\Omega_{r} =\Omega_{s}=15\gamma$. At high temperatures, inelastic collisions, resulting from mixing with other Rydberg states, also play a significant role~\cite{beigman1995Collision}. In high-density gases, and without loss of generality, we set $\gamma = 2\pi \times 19\,\text{MHz}$~\cite{bai2020SelfInduced, hu2023Improvement}. 
As sketched in Fig.~\ref{fig:4}, we present the Fourier-spectrum phase diagram as a function of the signal-field strength $E_{\mathrm{SIG}}$ for different detunings $\delta_s$.  The local field is fixed at $E_{\mathrm{LO}} = 7.85 \times 10^{-3} \, \mathrm{V/cm}$, with its locking point indicated by the red vertical line in panels (a1)--(c1). For a resonant signal field ($\delta_s = 0$), the CTC regime breaks down rapidly as the signal field increases. The oscillation frequency jumps from its CTC value to zero before the oscillation vanishes, bringing the system to equilibrium. The corresponding minimum detectable electric field strength under this condition is given by $E_{\mathrm{min}} = \Omega_{\mathrm{min}} \hbar / \mu \approx 0.6 \mathrm{nV/cm}$, where $\mu$ denotes the corresponding transition dipole moment [see the transition from point A to B in Fig.~\ref{fig:4}(a2)]. For non-zero detunings $\delta_s$ [Fig.~\ref{fig:4}(b1)-(b3) and (c1)-(c2)], we observe that the jump region broadens, and we obtain the minimum detectable field strength with $E_{\mathrm{min}} = 1.2\,\mathrm{nV/cm}$ and $8.4\,\mathrm{nV/cm}$ for $\delta_s=0.3\,\mathrm{MHz}$ and $1\,\mathrm{MHz}$, respectively. 
The observed increase in  $E_{\mathrm{min}}$ with detuning  $\delta_s$ is caused by off-resonant coupling. This match in behavior is consistent with the operation of a superheterodyne microwave receiver.
The limitation can be addressed by implementing a linear array of scalable Rydberg vapor cells equipped with a Stark comb~\cite{jiao2025Arbitrary}.
Therefore, these findings constitute a new CTC-based paradigm for MW detection in nonequilibrium systems.

\textit{Conclusion and discussion}.---
In this work, we have investigated dissipative phase transitions in a Rydberg many-body system driven by a tunable MW field. By exciting atoms into two-component Rydberg states, we have demonstrated the emergence of time crystals, with the oscillatory regime exhibiting exceptional sensitivity to MW perturbations.
A key finding is the observation of a rapid switching between distinct non-equilibrium states, that characterized by high- and zero- oscillation frequencies, at the boundary separating the monostable and oscillatory regimes as the MW field is tuned.
Furthermore, we have analyzed MW-controlled phase transitions in a Doppler-broadened medium. Our results have revealed that, due to the spontaneous breaking of time-translation symmetry and MF couplings, a significant fraction of atoms across different velocity classes exhibit synchronized oscillatory dynamics. 
By leveraging the phase transition near criticality in thermal gases, we have performed precision scans of the MW field. Theoretical calculations predict  a minimum detectable
MW field strength on the order of $1\,\text{nV/cm}$ in this many-body nonequilibrium system.
This study establishes a theoretical framework for understanding nonequilibrium phase transitions in MW-driven Rydberg systems, paving the way for future experiments and potential applications in quantum sensing and metrology.

\textbf{Acknowledgments}.---
We gratefully acknowledge Linjie Zhang, Yuechun Jiao, and Weibin Li for valuable discussions.
This work is supported by the National Natural Science Foundation of China  (12274131, 12347102, 12174184), Natural Science Foundation of Jiangsu Province (BK20233001), and Innovation Program for Quantum Science and Technology (2024ZD0300101).

\textbf{Conflict of interest.}---
The authors declare that they have no conflict of
interest. 

%
%
\end{document}